\documentclass[iop,apj]{emulateapj}
\usepackage{url}
\usepackage{natbib}
\shorttitle{Acoustic Frequencies of the Sun during Solar Cycle 23 and 24}
\shortauthors{Tripathy, Jain, and Hill}
\begin{document}

\title{Variations in High Degree Acoustic Mode Frequencies \\
    of the Sun during Solar Cycle 23 and 24}
\author{S.~C.~Tripathy, K.~Jain and F.~Hill}
\affil{National Solar Observatory, Tucson, AZ 85719, USA}
\email{stripathy@nso.edu}
\begin{abstract}
We examine continuous measurements of the high-degree acoustic mode frequencies of the Sun covering the period from 2001 July to June 2014.  These are obtained 
through the ring-diagram technique applied to the full-disk Doppler observations made by the Global Oscillation Network Group (GONG). The frequency shifts in the degree range of 180 $\le \ell \le 1200$ are  correlated with different proxies of solar activity e.g. 10.7~cm radio flux, the International Sunspot Number and the strength of the local magnetic field.  In general, a good agreement 
is found between the shifts and activity indices, and the correlation coefficients are found to be comparable with intermediate degree mode frequencies.  Analyzing the frequency shifts separately for the two cycles, we find that cycle 24 is weaker than cycle 23.  Since the magnetic activity is known to be different in the two hemisphere, for the first time, we compute the frequency shifts over the two hemispheres separately and find that the shifts also display hemispheric asymmetry; the amplitude of shifts in the northern hemisphere peaked  during late 2011,  more than two years earlier than the south.  We further correlate the hemispheric frequency shifts with  the hemispheric sunspot number and mean magnetic activity index.  Since the frequency shifts and the hemispheric activity indices are found to be significantly correlated, we suggest that the shifts be used as an indicator of  hemispheric activity  since not many indices are measured over the two hemispheres separately. We also investigate the variation at different latitudinal bands and conclude that  the shifts in active latitudes correlate well with the 
local magnetic activity index.   

\end{abstract}


\section{Introduction}\label{S-Introduction} 
The eigen-frequencies of solar oscillations which provide insight into the interior structure and dynamics of the Sun 
are known to vary with the solar activity \citep{sct-woodard85, sct-els90, sct-woodard91}. With improved observations, the relationship between  mode frequencies and activity cycle has been extensively studied and it is well 
established that the frequencies exhibit a good correlation with many different solar activity proxies 
\citep{sct-chap07, sct-jain09}. Comprehensive analyses using consistent and longer data sets both from ground- and space-based instruments, however, have shown different and unexpected results. For example, a phase-wise analysis of the intermediate degree frequencies
showed that the degree of correlation is different from phase to phase of the cycle \citep{sct-jain09}.  The extended minimum phase between cycle 23 and cycle 24 also revealed significant differences between acoustic frequencies and solar activity \citep{sct-bison09, sct-salabert09, sct-2010apj}. Analyzing Birmingham Solar Oscillations Network data over three solar cycles, \citet{sct-basu12} presented evidence that the solar sub-surface layer was very different between cycles 22 and 23, with deeper changes occurring during cycle 22 compared to cycle 23. A recent analysis of low-degree modes from the Global Oscillations at Low Frequency instrument also showed that cycle 24 is magnetically weaker in the upper sub-surface layers of the Sun \citep{sct-david15}.  

Since the frequency shifts are predominantly a function of the frequency alone, it has been suggested that the perturbation causing the changes are confined to the shallow layers \citep{sct-els94, sct-howe99, sct-chap01, sct-rosy13}. In addition, a small but statistically significant change in modes with turning points at or below the convection zone has also been reported \citep{sct-chou05, sct-baldner08}. Thus, the physical mechanism still remains an open question viz. whether the shifts are caused by structural changes in the sub-surface layers, such as changes in temperature \citep{sct-kuhn88} or  by the magnetic fields either through the action of the Lorentz force on the plasma or indirectly by modifying 
the physical properties e.g. size of the acoustic cavity \citep{sct-dzi05}. Another possibility is the influence of the magnetic atmosphere \citep{sct-rjain96}.   The analysis of frequency shifts by \citeauthor{sct-dzi05} further suggest that it is the indirect effect 
that dominate, in particular changes to the near surface stratification resulting from the suppression of convection by the magnetic field. On the 
other hand, they also reported small but significant departures from the simple scaling of the lower-frequency {\it p} mode frequency shifts with the inverse mode inertia implying that there is a contribution from deeper layers due to the direct effect of the magnetic fields. However, through theoretical modeling, \citet{sct-foullon05} concluded that the effect of a buried magnetic field on the frequencies of solar 
acoustic modes is too small to account for the changes in frequency over the solar cycle. Moreover, \citet{sct-howe02} showed that the temporal 
and latitudinal distribution of the frequency shifts is correlated with the spatial distribution of the surface magnetic field implying that the shifts cannot be purely explained by the structural changes. Thus, a thorough modeling is required to better understand the cause of the 
frequency shifts. 
 
\begin{figure}[h]    %%%%%%%%%%%%%%%%%% FIGURE 1
\plotone{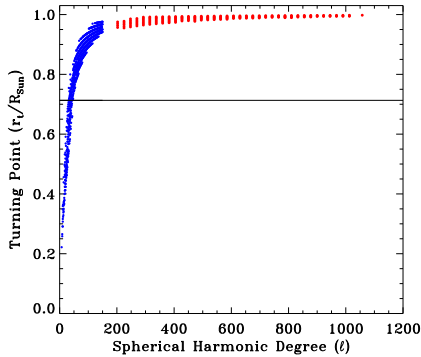}
              \caption{The lower turning point of oscillation modes for intermediate (blue circles) and high (red circles) as a function of the spherical harmonic degree, $\ell$. The spread is due to different radial orders.}
   \label{F-fig0}
   \end{figure}

In this context, it is desirable to analyze high-degree modes
whose lower turning points occur very close to the solar surface (Figure~\ref{F-fig0}). But determination of high-degree mode frequencies is a challenging task since 
individual modes blend into ridges at high-degree \citep[see, for example,]
[and references therein]{sct-kor2013}. In spite of  this inherent difficulty, there have been several attempts to compute and analyze high-degree oscillation frequencies. Using  time series from the Michelson Doppler Imager (MDI) on board the {\it Solar and Heliospheric Observatory} ({\it SOHO}),  \citet{sct-rhodes02} estimated the high-degree mode frequencies and demonstrated the response of the {\em p}-modes to short-term changes in solar activity.  \citet{sct-rhodes03} extended this analysis to include data from  Global Oscillation Network Group (GONG). Considering global modes from MDI, \citet{sct-crs08} concluded that the changes
in the high-degree mode frequencies are in good agreement with the intermediate-degree results, except for years when the MDI was highly defocussed. A similar study, but including more data, demonstrated a quadratic relation between the frequency shifts and activity indices \citep{sct-crs11}, which was not confirmed by \citet{sct-nso50}. \citet{sct-rhodes11} also investigated the temporal changes in high-degree {\em p}-mode frequencies and pointed out that short-duration, high-frequency shifts are more sensitive to changes in solar activity than the shifts measured by long-duration studies consisting only the lower frequencies. Recently, \citet{sct-kornso50} computed high-degree modes for three different epochs from three different instruments and compared the mode parameters.  

Ring-diagram analysis \citep{sct-hill88}, in addition to global helioseismic techniques,  
measures the characteristics of high-degree modes (180 $\le \ell \le 1200$) in localized regions and has been used primarily to investigate the variations 
around active regions \citep{sct-rajguru01, sct-rick02, sct-rab08}. \citet{sct-hindman00} found that the frequencies act as a tracer of local magnetic activity which was later shown to be less 
reliable during low activity periods \citep{sct11}. However, the response of the frequencies of the high-degree modes to  solar activity using a continuous and consistent data set 
has not been undertaken previously.  The results reported earlier mostly use discrete data sets obtained by the MDI instrument when it was operating in its high-resolution mode three months every year. Our goal, in this paper,  is to investigate the behavior of high degree mode frequencies over a period of about 13 years consisting of both the descending phase of cycle 23 and the ascending phase of cycle 24. This analysis supplements our previous study \citep[hereafter paper I]{sct-sol13} where we had primarily focused on the minimum phase of solar cycle 23. Since magnetic activity  (e.g. the  sunspot number and polar flux)  exhibit north-south hemispheric asymmetry, for the first time, we examine the variations in oscillation frequencies over these two hemispheres separately. We further explore the relationship between the frequency shifts and magnetic activity at different latitudes in both hemispheres. 

The rest of the paper is organized as follows. The data and the analysis procedure is described in $\S$~2, while the results are described in $\S$~3. Finally, we summarize the main conclusions in $\S$~4.

\section{Data Analysis} 
The Global Oscillation Network Group 
has been operating in high-resolution mode since July 2001 and produces full disk Doppler images of the Sun every minute. These images are then processed through the ring-diagram pipeline \citep{sct-corbard03} to produce the mode frequency tables.   In short, the ring-diagram technique \citep{sct-hill88} produces a three dimensional spectra (two in space, one in time) of a small region on the surface of the Sun tracked and remapped over a given time period.  In the GONG pipeline,  each  region (hereafter tile) covers an area of $15\degr$ $\times$ $15\degr$ in heliographic  longitude and latitude and is tracked for a period of 1664 minutes (hereafter ring-day). The centers of the tiles are separated by 7.5$\degr$  and a set of 189 overlapping tiles, referred to as dense-pack,
covers the region of $\pm$52.5\arcdeg\ in latitude and central meridian distance from the disk center.  The three dimensional spectra is  fitted with a Lorentz profile model \citep{sct-haber00} 
\begin{equation}
P(k_x, k_y, \nu) =  \frac{A}{(\nu - \nu_0 + k_xU_x +k_y U_y)^2 + \Gamma^2} + \frac{{b}}{k^3}, 
\end{equation}
where P is the oscillation power for a wave with a temporal frequency $\nu$ and the total wave number $k^2 = k_x^2 + k_y^2$. There are six parameters to be fitted: the zonal and meridional velocities, $U_x$ and $U_y$ respectively, the central frequency, $\nu$, the amplitude, $A$, the mode width, $\Gamma$ and 
the background power, $b$. For each tile, the peak fitting procedure returns about 200 modes  with degree, $\ell$, between 180 and 1200 and with frequency between 1700 and 5600 $\mu$Hz. It is, however, to be noted that these are not discrete modes in the same sense as those observed at lower degrees. The ring frequencies are taken from the GONG website\footnote{ \url{http://nisp.nso.edu}} and  cover the period from 2001 July 26 to 2014 June 19.

\begin{figure}[h]    %%%%%%%%%%%%%%%%%% FIGURE 2
\plotone{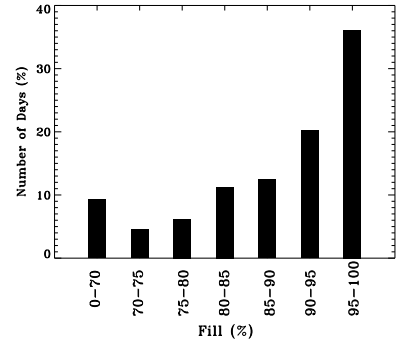}
              \caption{Fill factor of GONG ring-data between 2001 July 26 and 2014 June 19.}
   \label{F-fig1}
   \end{figure}

Since GONG is a ground-based network of six stations, poor observing conditions and instrument failures give rise to gaps in the data. These gaps are measured in terms of a fill factor, where a fill of unity means no data loss (100\% coverage). The distribution of the fill factor of the ring data are shown in Figure~\ref{F-fig1}.  Since gaps in the data causes biases and data of lower fill may not be fitted reliably, we exclude data where the fill factor is lower than 70\%.  This resulted in 3761 out of a total of 4146 ring-days and consists of approximately 710,800 tiles . 

The mode frequencies  (and other mode parameters) measured by the ring-diagram technique are affected by foreshortening as well as gaps in observation. Following the method outlined by \citet{sct-rhowe04}, we model the effects of position of the disk as a two-dimensional function of the distance from disk center $\rho$ without the cross-terms, combined with a linear dependence on the fill factor, $f(t)$,  

\begin{equation}
\nu(\rho_x,\rho_y,t)=a_0+a_1\rho_x+a_2\rho_x^2+a_3\rho_y+a_4\rho_y^2+a_5f(t), 
\end{equation}

\noindent where   $\rho_x$ and $\rho_y$ are the longitudinal and latitudinal component of 
 $\rho$ and $a_i$ are coefficients determined by fitting.   In order to minimize 
the effect of the magnetic field on the fitting, we use the frequencies over a two year period when solar activity is low (between 1 September 2007 and 31 August 2009)  and evaluated the $a_i$s for each ($n$, $\ell$) mode. The resultant coefficients and  the resultant differences between the corrected and original frequencies are shown in Figures~3 and 4 of paper I. 

\begin{figure}[t] %%%%%%%%%%%%%%%%%% FIGURE 3 
\plotone{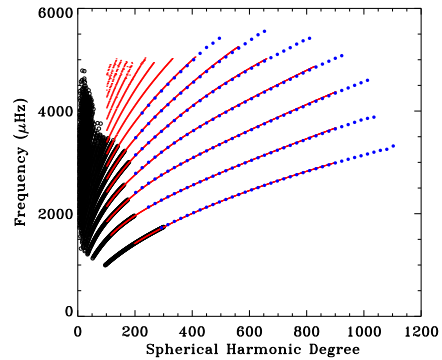}
\caption{The $\ell$--$\nu$ diagram of the modes obtained from the use of different techniques and data source. The black circles 
display the fitted intermediate degree modes using MDI data, the red dots represent the global high-degree modes from GONG, and the blue circles 
display the high-degree modes obtained through the ring-diagram analysis of the GONG data. \label{F-fig2}}
\end{figure}

Figure~\ref{F-fig2} compares the fitted modes obtained from global and ring-diagram analysis techniques in a characteristic $\ell-\nu$ plane. The black circles display the intermediate degree modes fitted using MDI data corresponding  to the epoch 2002 March 31 to 2002 June 10 \citep{sct-schou99}. The red circles represent the global high-degree modes from the GONG data obtained\footnote{\url
{https://www.cfa.harvard.edu/$\backsim$sylvain/research/tables/HiL}} from a time series of  98 days in length with a start date of 2002 February 14 \citep{sct-kornso50}.  The blue circles 
display the high-degree modes obtained by the ring-diagram analysis averaged over the same 98 days; the lowest ridge corresponds to the {\em f} mode while others are the {\em p}-mode ridges for radial order $n$ from 1 to 6. It is evident that the frequencies computed  from different analysis mostly agree with each other, as the symbols appear to fall along the same ridges. This also suggests that the systematic errors in the frequencies 
arising from analysis choices are independent of time. 

\begin{figure}[ht]    %%%%%%%%%%%%%%%%%% FIGURES 4 
   \plotone{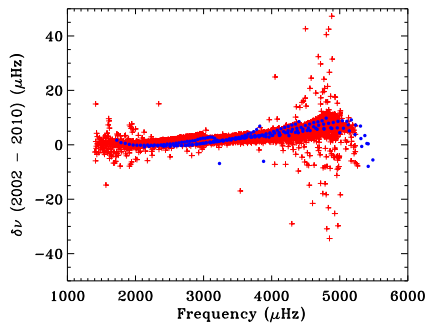}
    \caption{Frequency differences between 2002 and 2010. The plus symbols denote the differences computed from high-degree global modes obtained from Korzennik's fitting in the degree range of 100 and 1000. The filled circles denote the differences between high degree modes computed using the ring-diagram technique. The 2002 frequencies are calculated from a time series of 98 days with a start day of 2002 February 14, while those in 2010 are computed from a time series of 67 days starting from 2010 May 7.}
   \label{F-fig3}
   \end{figure}

We further compare the individual frequencies obtained from GONG data between high- (2002) and low- (2010) activity periods where the frequencies corresponding to low activity period are computed from a 67 day time series starting from 2010 May 7 \citep{sct-kornso50}. Figure~\ref{F-fig3} shows the frequency differences. The pluses denote the differences between the global high-degree modes  while the filled circles shows the differences between the modes obtained from the ring-diagram analysis.  The mode frequencies obtained from two different methods agree very well except for the scatter seen in the global modes at high frequencies. 

 \begin{figure}[h]    %%%%%%%%%%%%%%%%%% FIGURES 5
\plotone{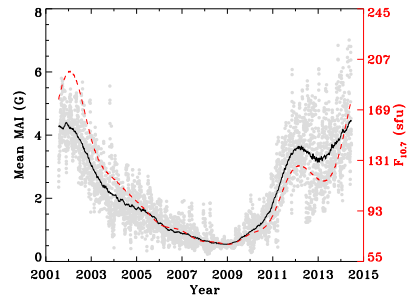}
    \caption{Temporal evolution of the mean magnetic activity index, MMAI. The solid black line shows the running mean over 181 points. The dashed red line represents the 181 day smoothed 10.7 cm radio flux measurements scaled linearly with MMAI.
   \label{F-fig4}}
   \end{figure}  
 
For the purpose of characterizing activity associated with each tile, we compute a magnetic activity index (MAI) which indicates the local magnetic field strength associated with each tile. The index is calculated by tracking and remapping the line-of-sight magnetograms during the same interval as the Dopplergrams  and averaging the absolute values of all the pixels with a field strength higher than a threshold value.  Since the entire period under study is not covered by the MDI magnetograms alone, we use the 96 minute MDI magnetograms up to 2007 April 9 and GONG magnetograms sampled every 32 minutes thereafter.  A lower threshold value of 50~G and 8.8~G are  used for MDI \citep{sct-basu04} and GONG respectively, which are approximately 2.5 times the estimated noise in individual magnetograms. For generating a uniform magnetic activity index, the MDI MAI values are scaled to GONG values using the conversion factor of 0.31 that  was obtained by relating 
the line-of-sight magnetic field measurements between these two instruments using a histogram-equalizing technique \citep{sct-harvey14}. 

A well known systematic effect in solar physics and specially with ground-based resolved data is the annual variation with the solar inclination toward Earth which can lead to a loss in spatial resolution and hence change the geometric foreshortening. This effect is clearly seen in the MAI, the frequency shifts (calculated later) as well as in solar activity indices. We have applied a technique based on Fourier Transform to remove the 1 year 
periodicity from all the date sets used in the study of temporal variations. However, for a better demonstration, uncorrected data has been included in some figures by means of symbols. Also note that the frequency shifts and MAIs have been interpolated to 4146 ring-days before the Fourier Transform was applied.  

 Figure~\ref{F-fig4} shows the temporal variation of the mean MAIs (MMAI) where the mean is taken over 189 dense-pack tiles of each ring-day. We also calculate a 181 point running mean, which corresponds approximately to seven Carrington rotations in terms of ring-days, to smooth out the short-term variations. This is shown by solid (or dashed) lines in most of the figures. In the same plot, we also show the variation of the 10.7~cm radio flux, $F_{10.7}$ as a measure of the solar activity (dashed red line).  Significant agreement is seen  between the shape of the curves  indicating a high correlation which is confirmed with a rank correlation coefficient value of 0.97. 
Thus the MMAIs will be considered as an activity proxy in rest of the  analysis.   

In order to examine the sensitivity of the frequency shifts to solar activity, we further use two  commonly used activity proxies; the International Sunspot Number, ISN, and $F_{10.7}$ both obtained from National Geophysical Data Center (NGDC)\footnote{\url{http://www.ngdc.noaa.gov/stp/solar/solardataservices.html}}.  The sunspot number is a weighted estimate of the individual sunspots and sunspot groups derived from inspection of the solar photosphere in white light while the radio flux index is a measurement of the integrated emission at 10.7~cm wavelength from all sources present on the solar disk.

\section{Results and Discussions}\label{S-result} 

\subsection{Spatial Variation}
\citet{sct-hindman00} studied the spatial variation of the frequencies obtained from ring-diagram analysis and showed that the active regions appear 
as locations of large frequency shifts. This gave rise to the conclusion that the frequency shifts can be used as a tracer of surface magnetic activity. Subsequent analysis \citep{sct10iia,sct11}, however, indicated that the conclusion is valid only during the period of high activity and breaks down during the activity minimum period. 

We estimate the spatial frequency shift, $\delta\nu_s$, weighted by the fitting error, $\sigma_{n,l}$, for each tile according to  
\begin{equation}
\delta\nu_s(t)\,=\,\sum_{n,\ell}\frac{\delta\nu_{n,\ell}(t)}{\sigma_{n,\ell}^2}
/\sum_{n,\ell}\frac{1}{\sigma_{n,\ell}^2} , 
\end{equation}
where the frequency difference of each mode, $\delta\nu_{n,\ell}$, is calculated with respect to the error weighted average of that particular mode 
present in the dense-pack of each ring-day. A sequence of images showing $\delta\nu_s$ and MAI for four days during the high- and low-activity periods are presented in Figures~1 and~2 of \citet{sct11} demonstrating the relationship between them.  

\begin{figure*}    %%%%%%%%%%%%%%%%%% FIGURE 6 
\plotone{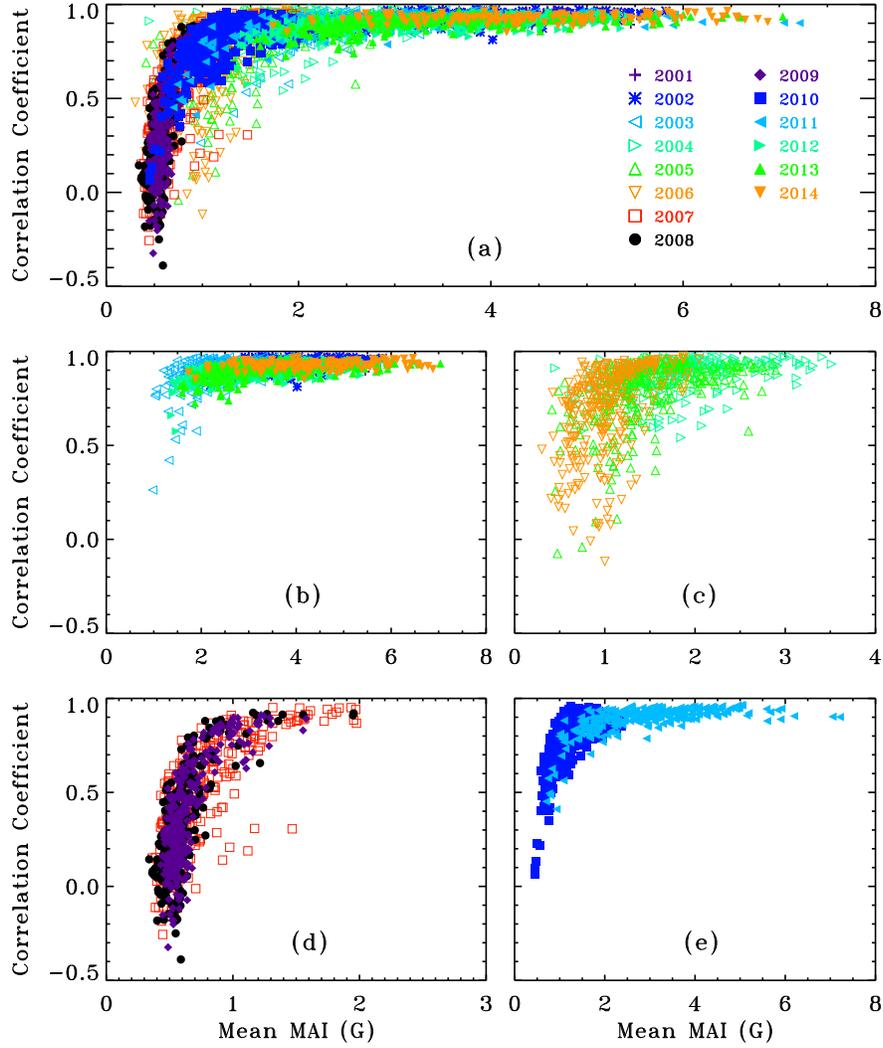}  
%\plotone{fig6b.ps}  
              \caption{(a) Pearson's linear correlation coefficient, $r_p$, between $\delta\nu_s$ and MAI over 189 dense-pack tiles for each ring-day as a function of the mean MAI. The symbols and colors represent the year-wise variation and are indicated in the top panel. Panels (b-e) shows the coefficients around the maximum (2001--2003; 2012--2014), descending (2004-2006), minimum (2007--2009) and ascending (2010--2011) phases of the solar activity in cycles 23 and 24, respectively. Note that the range of Mean MAI in each panel is different.}
   \label{F-fig5}
   \end{figure*}

For a quantitative analysis, we calculate the Pearson's linear correlation coefficient, $r_p$, between $\delta\nu_s$ and MAI of each tile in the dense-pack. The resultant coefficients (color coded for different years) as a function of the MMAIs are shown in Figure~\ref{F-fig5}a.
We observe that for high values of MMAI the correlation is stronger while for MMAI  $\le$ 1~G, the correlation coefficients 
are small with a median value of 0.41.  On closer inspection, we also find higher correlation coefficients for small MMAI's specially during the minimum phase. For a better visualization, we plot the correlation coefficients for different phases separately. Figure~\ref{F-fig5}b shows the period around the maximum phase of cycles 23 (2001--2003) and 24 (2012--2014) and illustrates stronger correlation between the frequency shifts and mean  magnetic activity index.  The coefficients for the descending phase of cycle 23 (2004--2006), the extended minimum phase (2007--2009), and 
the ascending phase of cycle 24 (2010--2011)  are shown in panels c, d and e of Figure~\ref{F-fig5}, respectively. In all cases, we note both high and low correlation coefficients for MMAI values of $\approx$ 1~G  and a large spread
in the values of the coefficients (largest for the extended minimum phase). We further observe negative coefficients in panels c and d (fewer in panel d) implying anti-correlation but the values are not significant. Analysis of global mode 
frequencies with solar activity proxies during the extended minimum phase between cycles 23 and 24 has also indicated anti-correlation between them   \citep{sct-salabert09,sct-2010apj}. Thus our result is consistent with the global low- and intermediate- degree modes.   

\begin{figure}    %%%%%%%%%%%%%%%%%% FIGURE 6 
\plotone{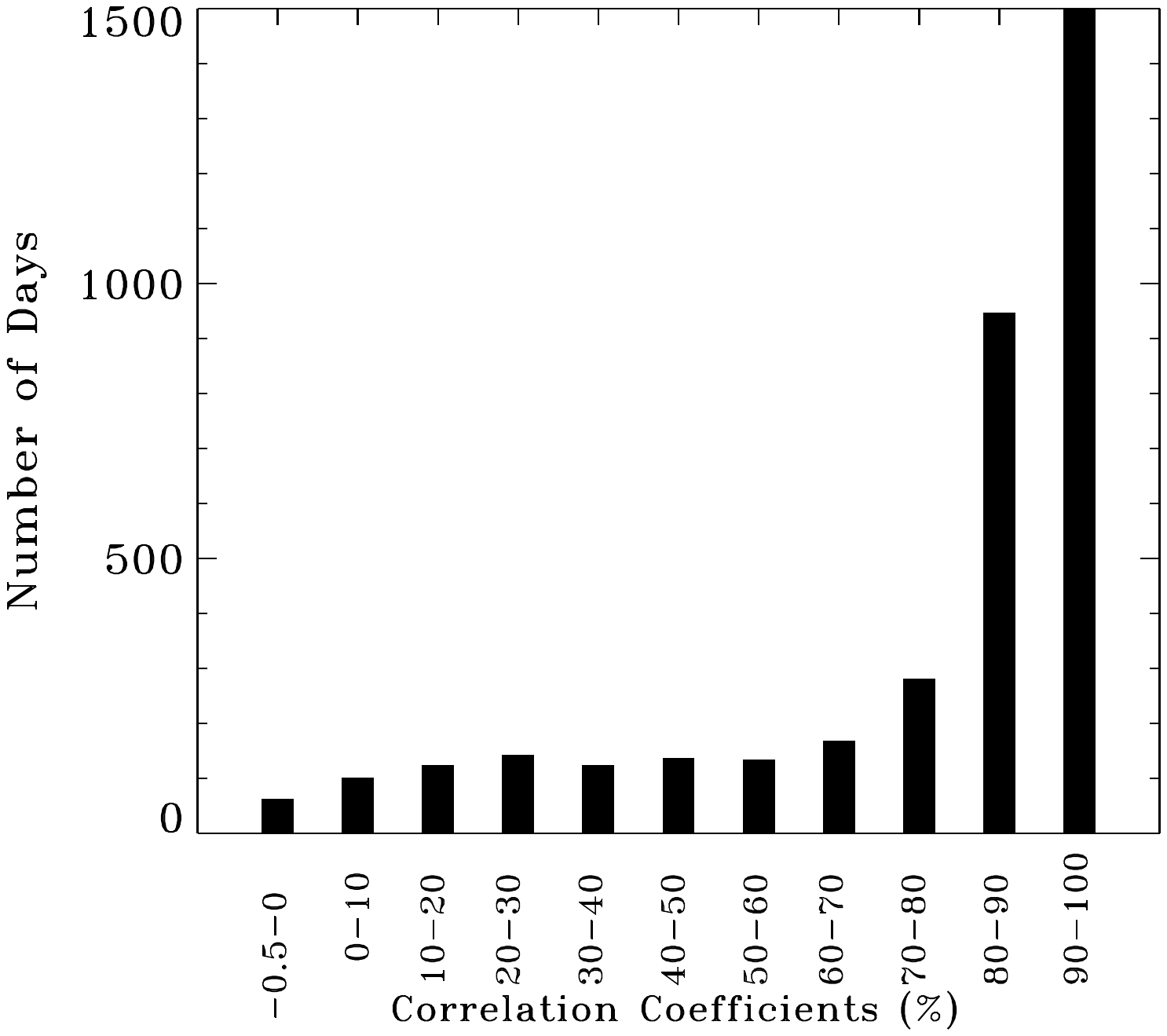}  
\caption{Histogram showing the distribution of the correlation coefficients as a function of 
the number of days.}\label{F-fig5b}
\end{figure}

The distribution of the linear correlation coefficient in the form of a histogram is shown in Figure~\ref{F-fig5b}. Only 
a small fraction of  63 ring-days (1.7\%)  are anti-correlated with MMAI while 73\% and 65\% ring-days have positive correlation coefficients higher than 70\%  and 80\%, respectively. 

\begin{figure*}    %%%%%%%%%%%%%%%%%% FIGURE 7 
   \plotone{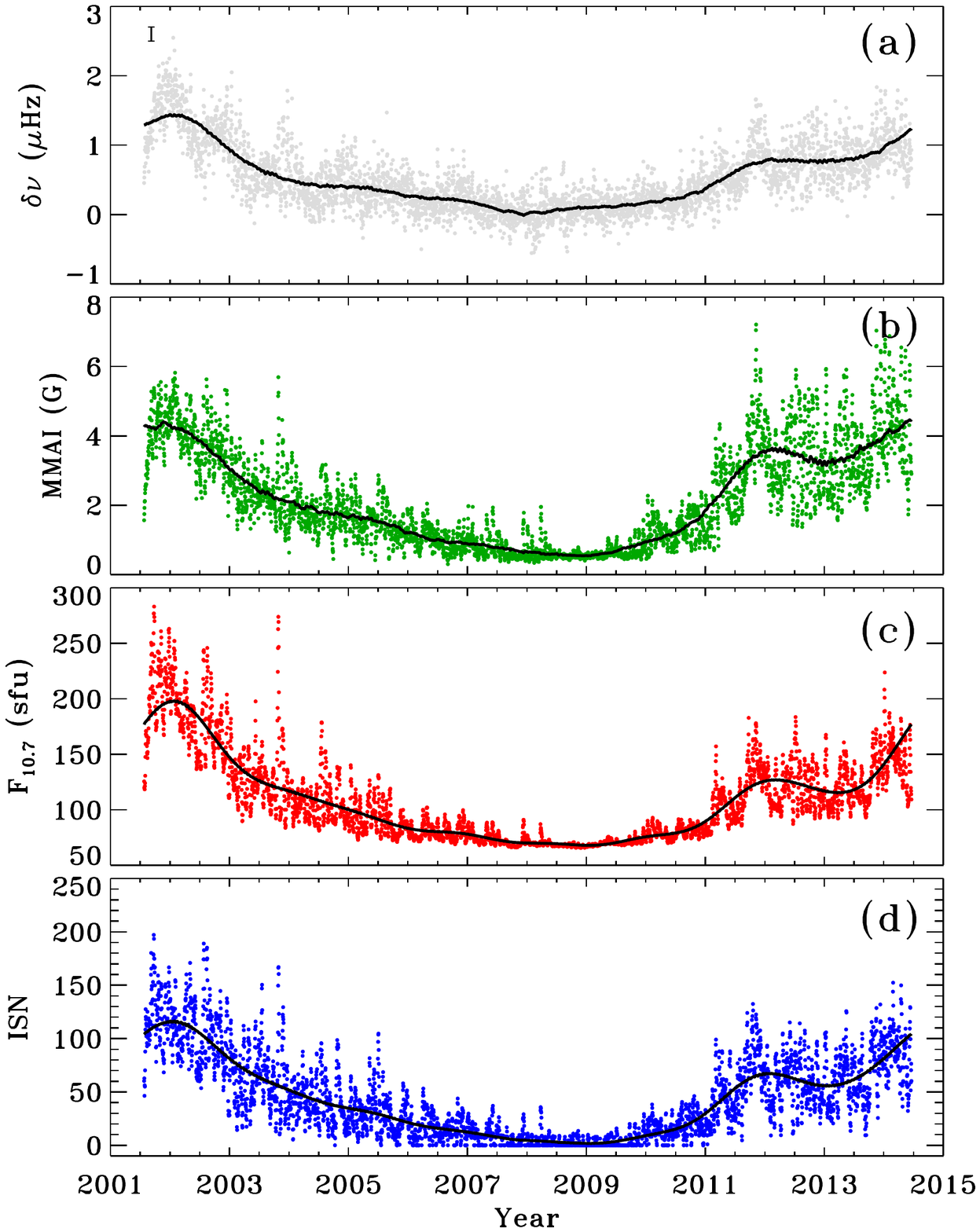}
   \caption{Temporal evolution of (a) $\delta\nu$, (b) MMAI, (c) F$_{\rm 10.7}$  and (d) ISN. The solid black line in each panel represents the running mean over 181 points. The error bar on the top left side of panel (a) represents 20 times the mean error. }
   \label{F-fig6}
   \end{figure*}

\begin{figure*}    %%%%%%%%%%%%%%%%%% FIGURE 8 
   \plotone{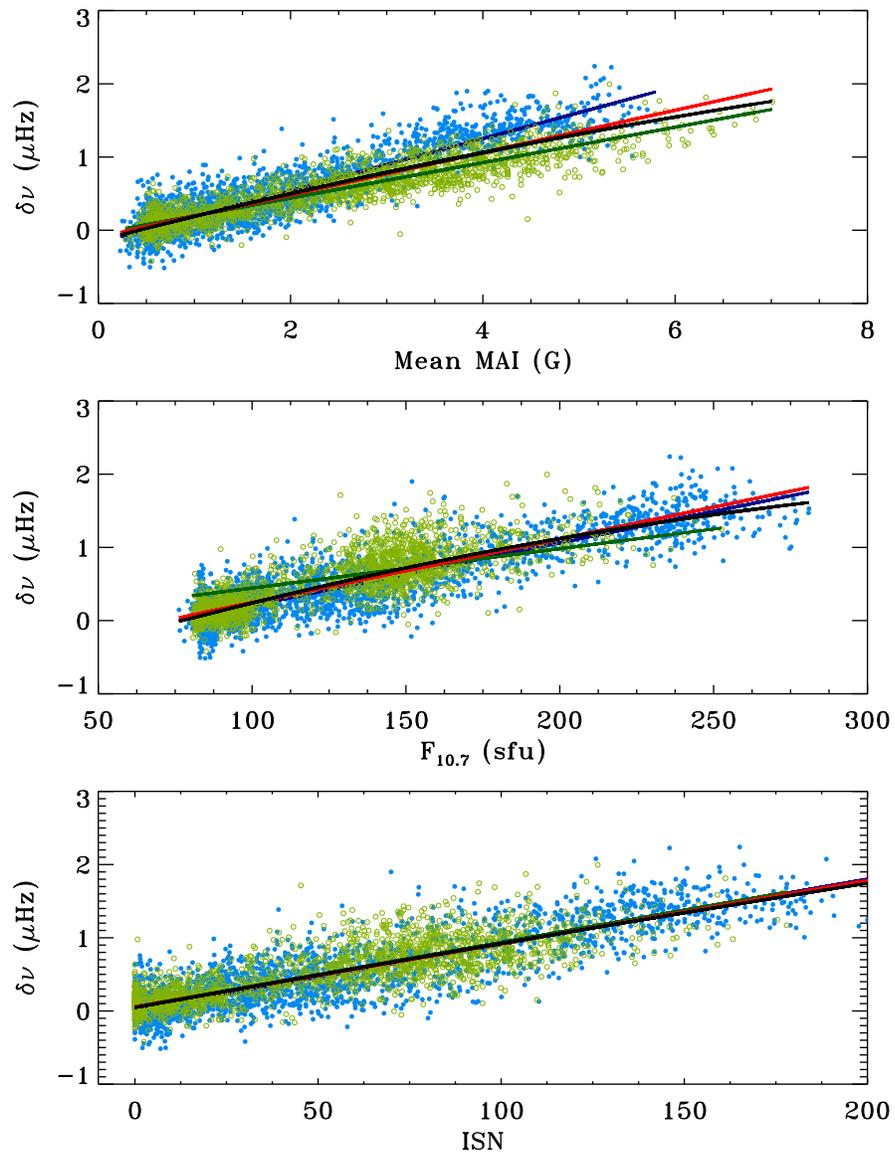}
   \caption{Correlations between the frequency shifts and different activity indicators for cycle 23 descending (filled circles) and 
   cycle 24 ascending (open circles) phases. The straight lines are the linear regressions for cycle 23 (blue), cycle 24 (green) and all data (red). The activity proxies are given on the abscissae. The black line is the result of a quadratic fit using all data.}
   \label{F-fig7}
   \end{figure*}

\subsection{Variation with Solar Cycle}\label{S-temporal}
The frequency variation with the solar cycle is investigated by calculating a mean frequency shift, $\delta\nu$,  
using a formula which is generally applied to compute frequency shifts of global modes, 
\begin{equation}
\delta\nu\;=\,\sum_{n,\ell}\frac{\sum_{i=1}^{189}\,\delta\nu_{n,\ell}(t)}{\sigma_{n,\ell}^2}
/\sum_{n,\ell}\frac{1}{\sigma_{n,\ell}^2}, 
\end{equation}
where $\delta\nu_{n,\ell}$ is  the frequency difference of each ($n$, $\ell$) mode with respect to the average frequency of the same mode over the dense-pack tiles corresponding to a quiet ring-day 2011 May 2008.  Table~\ref{T-tab0} shows the values attained by the mean frequency shifts and three activity proxies during cycles 23 (up to 2008 December 31) and 24. However, the values presented here may not be the true minimum and maximum values of the solar cycles (and hence the ratio) due to the period considered in this study. The final column in the table shows the ratio of amplitudes of cycle 23 to cycle 24, in the sense (max--min)$_{23}$/(max--min)$_{24}$. While the mean shifts and activity indices show a higher amplitude in cycle 23, the MMAI, which is the strong field component 
of the magnetic flux, shows a higher amplitude in cycle 24.  This is also evident   in Figure~\ref{F-fig6}, where $\delta\nu$ and activity proxies 
are plotted as a function of time. The solid line in each panel represents the  running mean over 181 points and guides the eye to visualize the descending  phase of cycle 23, ascending phase of cycle 24 and the extended minimum period between the two cycles. While the extended minimum phase is unambiguously  detectable in the activity indices, it is not noticeable in the frequency shifts illustrating that (i) the change in frequency with activity is no longer linear
during the solar minimum and (ii) the minimum phase in frequency lasted over a shorter period compared to the solar activity.  This is consistent with our earlier conjecture that the changes in high-degree frequencies  may be associated with two components: strongly localized active regions and the small scale horizontal magnetic field \citep{sct10iia}. The former explains the larger shifts, while the later is responsible for the smaller shifts
in all phases of the solar activity including the minimum phase. We return to this discussion in $\S$~\ref{S-min}.

We further examine the sensitivity of the frequency shifts to solar activity by performing a linear fit between them using data over the entire 13 year period as well over different phases separately (Figure~\ref{F-fig7}). The straight lines are the result of linear regressions for cycle 23 (blue), cycle 24 (green) and all data (red). The black solid line in each of the panels is the result of a quadratic fit using all data and differs  from the linear fit (red line) when activity is high.  Thus, we conclude that the activity proxies display simple linear relations with the frequency shifts instead of a quadratic relation as reported earlier \citep{sct-crs11}.  We conjecture that the quadratic relation is an outcome 
of using fewer data points as we observe at high-activity period.   Table~\ref{T-tab1} shows the frequency shift per unit activity gradients and the corresponding linear correlation coefficients for each of the proxies. For MMAI, the gradient is higher for cycle 23 compared to cycle 24, while for other 
proxies the gradients do not vary significantly between different cycles or over the entire time span.  The level of correlation with different 
activity proxies are higher than 80\% largely in good agreement with the analysis of intermediate degree global modes \citep{sct-jain11};  
there are, however, marginal changes between cycle 23 and 24 and entire period. Also, as anticipated, the correlation is  higher between $\delta\nu$ and magnetic activity measured locally. 

Before the data were corrected for the annual variation,  a  secondary periodicity except during the minimum phase was observed in the frequency shifts as well as in activity indices.    This signal is attributed to the quasi-biennial signal \citep{sct-anne12, sct-rosy13} that has been recently received a great deal of interest but is beyond the scope of this paper. 
 
\begin{deluxetable*}{lcccccc}
\tablewidth{0pt}
\tabletypesize{\scriptsize} 
\tablecaption{Values of Frequency Shifts and Activity Proxies for Different Solar Cycles\label{T-tab0}}
\tablehead{\colhead{}&\multicolumn{2}{c}{Cycle 23} &\colhead{}&\multicolumn{2}{c}{Cycle 24}&\colhead{}\\
\cline{2-3} \cline{5-6}
\colhead{Activity Index}& \colhead{Minimum}& \colhead{Maximum}&\colhead{}&\colhead{Minimum}& \colhead{Maximum}&\colhead{(Max -- Min)$_{23}$/(Max -- Min)$_{24}$}}
\startdata
$\delta\nu$&$-$ 0.52 & 2.24&&$-$ 0.42&1.99&1.14\phn\\
MMAI&\phn{0.23}&5.80&&0.28&7.01&0.83\phn\\
F$_{10.7}$&76.2&281&&80.4&253&1.19\phn\\
ISN&{0}&200&&0&178&1.12\phn
\enddata
\tablecomments{The MMAI has units of Gauss; the 10.7 cm radio flux is expressed in sfu (1 sfu = 10$^{-22}$ J s$^{-1}$
m$^{-2}$ Hz$^{-1}$); ISN is dimensionless and $\delta\nu$ is in $\mu$Hz.} 
\end{deluxetable*}

\subsection{Hemispheric and Latitudinal Variation \label{S-latitude}}
In addition to the temporal changes, magnetic activity is generally north-south asymmetric and varies with latitudes. The hemispheric asymmetry in  sunspot number and polar flux is well established and provides important clues for understanding the mechanism of solar cycle \citep [for 
a review see,][]{sct-aanorton2014}. The variation can be easily seen in magnetic butterfly diagrams where the activity belts move towards the geometric equator with the progression of the solar cycle. In addition, the wings of the butterfly diagram show a high degree of mirroring across the equator indicating coupling between the two hemispheres.    In this context, 
the frequencies obtained from ring-diagram technique provides a natural way to investigate variations between different  hemispheres  as well as latitudes since the $m$-averaged global frequencies of oscillation modes, commonly used for solar cycle studies,  are latitudinally averaged and can not distinguish between the northern and southern hemispheres. Nonetheless, the frequency of individual multiplets (which are function of $n$, $\ell$, $m$) 
can be be used to study  latitudinal variations based on $m$/$\ell$ values \citep[e.g.][]{sct-howe02, sct-jain11}. The sectoral and near-sectoral modes are sensitive to the region near the equator while the zonal and tesseral modes are sensitive to the wider range of latitudes.

\begin{figure*}                              % FIG 9 
\plotone{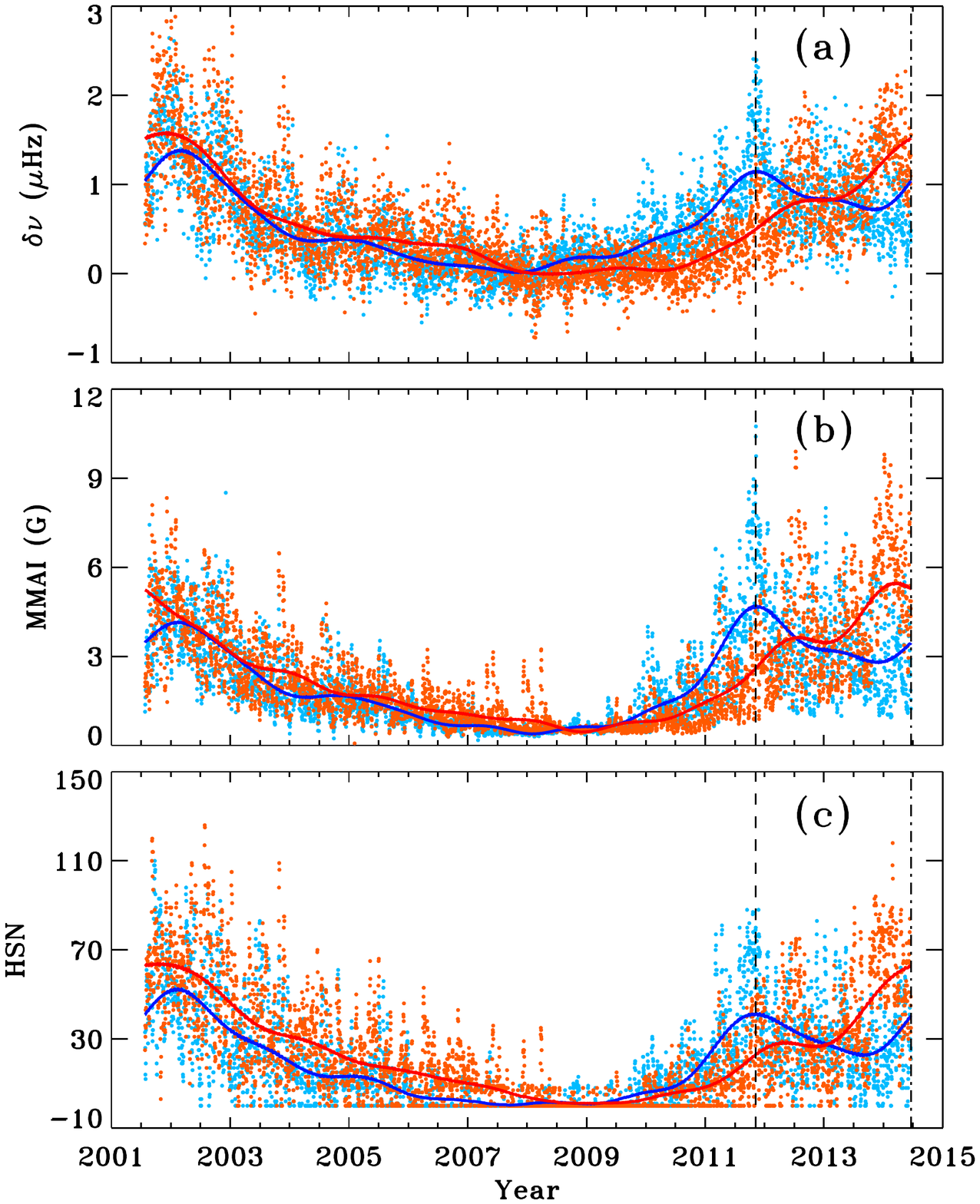}
              \caption{Temporal variation of (a) $\delta\nu$, (b) magnetic activity and (c) hemispheric sunspot number (HSN) for northern hemisphere (blue circles) and  southern hemisphere (orange circles).  The solid line in each panel represents the running mean over 181 points, the blue and red lines 
denoting northern and southern hemispheres, respectively. The dashed and dot-dashed lines indicated the peak amplitudes in northern and southern hemispheres, respectively. }
\label{F-fig8}
   \end{figure*}

\begin{deluxetable*}{llcclcclc}
\tablewidth{0pt}
\tabletypesize{\scriptsize} 
\tablecaption{Gradients and Linear Correlation Coefficients of Frequency Shifts as a Function of Activity Proxies \label{T-tab1}}

\tablehead{\colhead{}&\multicolumn{2}{c}{Cycle 23} &\colhead{}&\multicolumn{2}{c}{Cycle 24} &\colhead{}&\multicolumn{2}{c}{All Data}\\
\cline{2-3} \cline{5-6} \cline{8-9}\\
 \colhead{Activity Index}& \colhead{Gradient}&\colhead{$r_p$}&\colhead{}& \colhead{Gradient}&\colhead{$r_p$}&\colhead{}&\colhead{Gradient}&\colhead{$r_p$}
}
\startdata
MMAI&0.355 $\pm$ 1.58$\times 10^{-4}$ &0.91&&0.242 $\pm$ 1.60$\times 10^{-4}$&0.90&&0.289 $\pm$ 1.09$\times 10^{-1}$&0.88\\
F$_{10.7}$ &0.009 $\pm$ 3.99$\times 10^{-6}$&0.87&&0.005 $\pm$ 3.14$\times 10^{-6}$&0.79&&0.009 $\pm$ 3.45$\times 10^{-3}$&0.87\\
ISN&0.009 $\pm$ 4.22$\times 10^{-6}$&0.87&& 0.009 $\pm$ 4.89$\times 10^{-6}$&0.79&&0.009 $\pm$ 3.45$\times 10^{-3}$& 0.87
 \enddata
\tablecomments{Units of the gradients are $\mu$Hz per unit activity, where the activity units are defined in Table~\ref{T-tab0}.}
\end{deluxetable*} 
 Figure~\ref{F-fig8} shows the mean frequency shifts for  northern (blue circles) and southern (orange circles) hemispheres and the solid line represents the running mean over 181 points. The shifts are calculated using Equation~(3) where $\delta\nu_{n,l}$ corresponds to the  frequency difference of each mode with respect to  an average frequency over the same hemisphere. Both the northern and southern frequency shifts decreased starting from mid-2001 and their variations are similar until the end of 2007. Around the beginning of the cycle 24, the two shifts diverged and reached  their maximum amplitude at two different periods. In cycle 24, the northern shifts peaked in late 2011, more than two years earlier than the south, as indicated by the dashed and dot-dashed lines, respectively. This closely agrees with the hemispheric asymmetry seen in other activity indices e.g. the MMAI (panel b) and the hemispheric sunspot number (HSN, panel c), where the latter is obtained from the NGDC. Conspicuously, both the dashed and dot-dashed lines are aligned between different panels, illustrating that the frequency shifts very closely follow the changes in activity indices over both hemispheres.   
To quantify the relationship, we compute  the correlation coefficient between them and the shift per unit activity gradients. These are tabulated in Table~\ref{T-tab2}. The relative change in the gradients points out that the shifts  remained approximately constant between the two hemispheres. Moreover, the Pearson's linear and Spearman's rank correlation coefficients further confirms 
the linear relation between the shifts and activity indices. The values mentioned in the brackets are the coefficients corresponding to the smoothed values and, as expected, signify a closer interaction. In the absence of many activity proxies for different hemispheres, we suggest that the frequency shifts may be used as an hemispheric activity proxy.  
\begin{figure*}                      % Fig 10        
\plotone{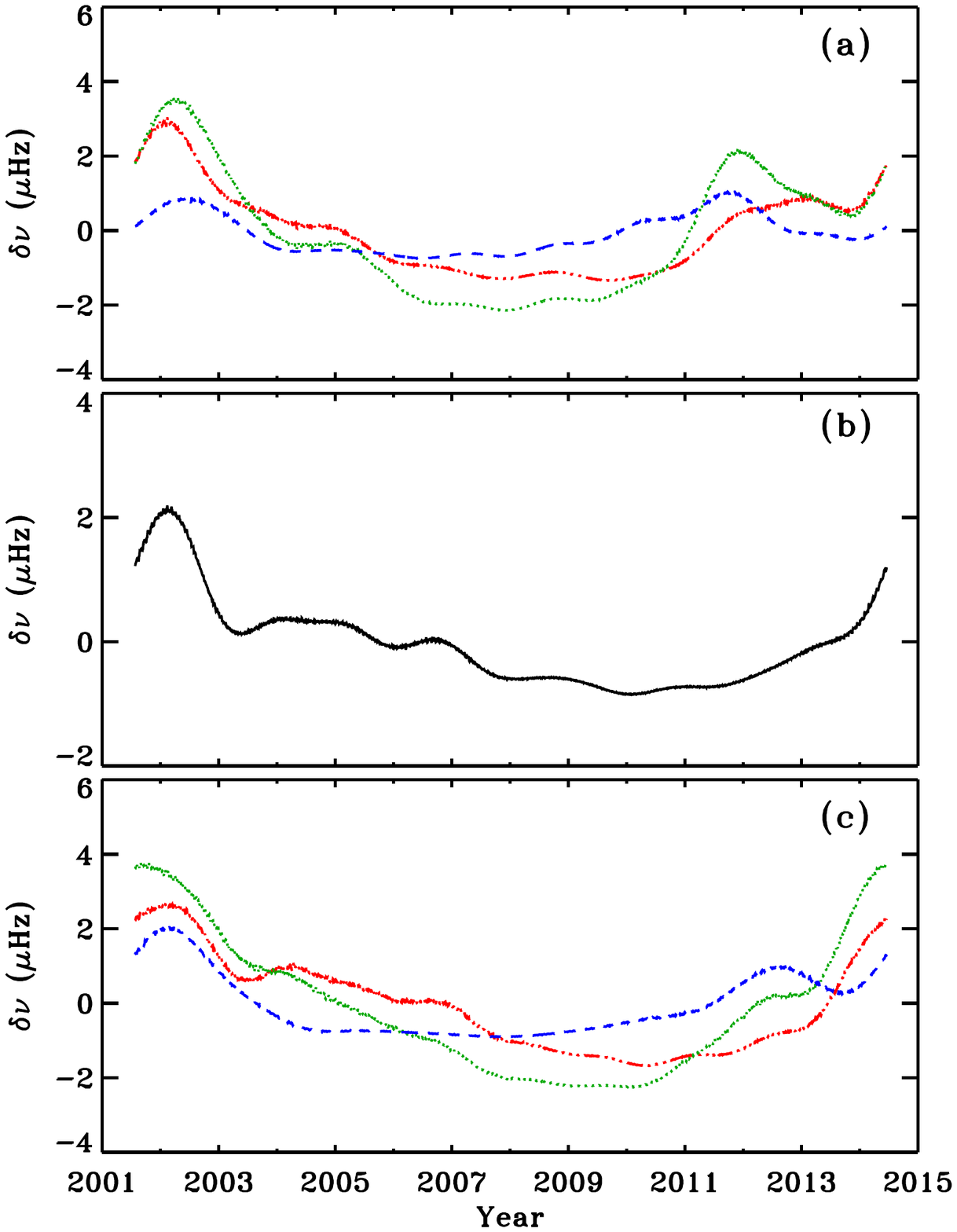}
              \caption{Temporal variation of frequency shifts as a running mean over 181 points at selected latitudes on northern (panel a) and  southern (panel c) hemispheres.   Line styles have the following meaning: red (dash-dot-dot-dot): 7.5\arcdeg, green (dotted): 15\arcdeg, and blue (dashed): 30\arcdeg.  Panel (b) shows the shifts corresponding to the equatorial tiles.\label{F-fig9}}
   \end{figure*}

  \begin{deluxetable*}{llcccc}
\tablewidth{0pt}
\tabletypesize{\scriptsize} 
\tablecaption{Values of Activity Proxies, Frequency Shifts and Correlation Statistics between $\delta\nu$ and Activity Indices (Pearson Linear Correlation, $r_p$, and Spearman rank Correlation, $r_s$). \label{T-tab2}}
\tablehead{
\colhead{Activity Index}& \colhead{Min}& \colhead{Max}&\colhead{Gradient}&\colhead{$r_p$}&\colhead{$r_s$\tablenotemark{a}}}
\startdata
&&&Northern Hemisphere&\\
\cline{4-4}
&&&&\\
MMAI (G) &0.04&8.09&0.285 $\pm$ 1.61$\times 10^{-4}$ &0.90 (0.97)\tablenotemark{b}&0.90 (0.97)\\
HSN&0&110&0.022 $\pm$ 1.32$\times 10^{-5}$&0.84 (0.98) &0.84 (0.97)\\
%$\delta\nu$&$-$ 0.640&2.65&\nodata&\nodata&\nodata\\
&&&&\\
&&&Southern Hemisphere&\\
\cline{4-4}
&&&&\\
MMAI (G) &0.00&0.00&0.284 $\pm$ 1.34$\times 10^{-4}$&0.86 (0.94)&0.87 (0.95)\\
HSN&0.0&126&0.022 $\pm$ 1.09$\times 10^{-5}$&0.86 (0.98)&0.85 (0.98)\\
%$\delta\nu$&$-$ 0.720&2.88&\nodata&\nodata&\nodata
\enddata
\tablenotetext{a}{The calculated probabilities of having null correlations are 0.}
\tablenotetext{b}{The values in the brackets are the correlation coefficients corresponding to the 181 point running mean.}
\end{deluxetable*}

 Figure~\ref{F-fig9} shows the mean frequency shifts 
for a few selected latitudes where the shifts are obtained by subtracting the average frequency computed at the same latitude covering all data sets but only over the central meridian. For clarity, we only show the mean shifts resulting from a running mean over 181 points. The shifts show significant variations between different latitudes as well as between the two cycles. Moreover, a clear extended minimum phase is not seen at any of the latitudes. In cycle 23, peak amplitudes are observed around the beginning of 2002 for latitudes 7.5\arcdeg~N and 15\arcdeg~N, while for southern hemisphere the  peak amplitude can be seen for all the latitudes around the same time. On the contrary,  we notice opposite behavior in cycle 24; the peak amplitude is seen around the end of 2011  for all latitudes while in the southern hemisphere, the shifts peak around the beginning of 2014 except for 30\arcdeg\   
latitude. At this  latitude, a peak amplitude is observed around 2012 November and a decreasing trend afterwords. Interestingly, the shifts corresponding to the equatorial belt has only one strong peak amplitude a 2002 and a highly reduced amplitude (about half of cycle 23) for cycle 24.  Thus, if we consider frequency shifts that correspond to equator we will conclude that cycle 24 is weaker compared to cycle 23 while at other latitudes, the shifts 
are nearly similar (slightly higher for southern hemisphere) between cycles 23 and 24.      

 \begin{figure}                              % FIG 11 
\plotone{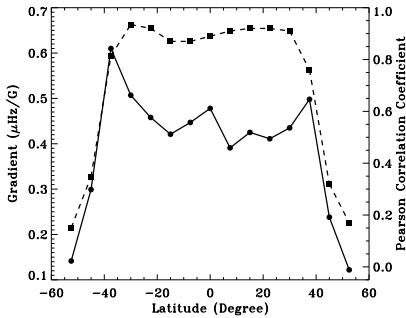}
              \caption{Variation of linear regression slopes (circles) and Pearson's correlation coefficients (squares) for different latitudes.}
\label{F-fig11}
   \end{figure} 

The frequency shift per unit magnetic activity (filled circles) and the Pearson's  linear correlation coefficients (filled squares) 
between MAI and $\delta\nu$ for different latitudes are shown in Figure~\ref{F-fig11}. Higher values of the gradients are found corresponding to the solar active latitude belt i.e. about $\pm$ 30\arcdeg. In the same latitudes, magnetic activity as measured by MAI are also found to be larger and hence a strong correlation is observed between the MAI and $\delta\nu$. The correlation coefficient decreases to lower than 0.8  
at latitudes higher than $\pm$ 37.5\arcdeg\ where the presence of the active region is sparse. From the nature of the two curves, it appears that the linear regression slopes and the magnitude 
of the correlation are well related. A similar result has been reported for frequency shifts in global modes \citep{sct-sol07}.  
For a robust comparison, it will be worth analyzing the entire solar cycle 24 in terms of the ascending and descending phases when the data is available. 

\subsection{Minimum Phase}\label{S-min}
\begin{figure*}    %%%%%%%%%%%%%%%%%% FIGURE 8 
   \plotone{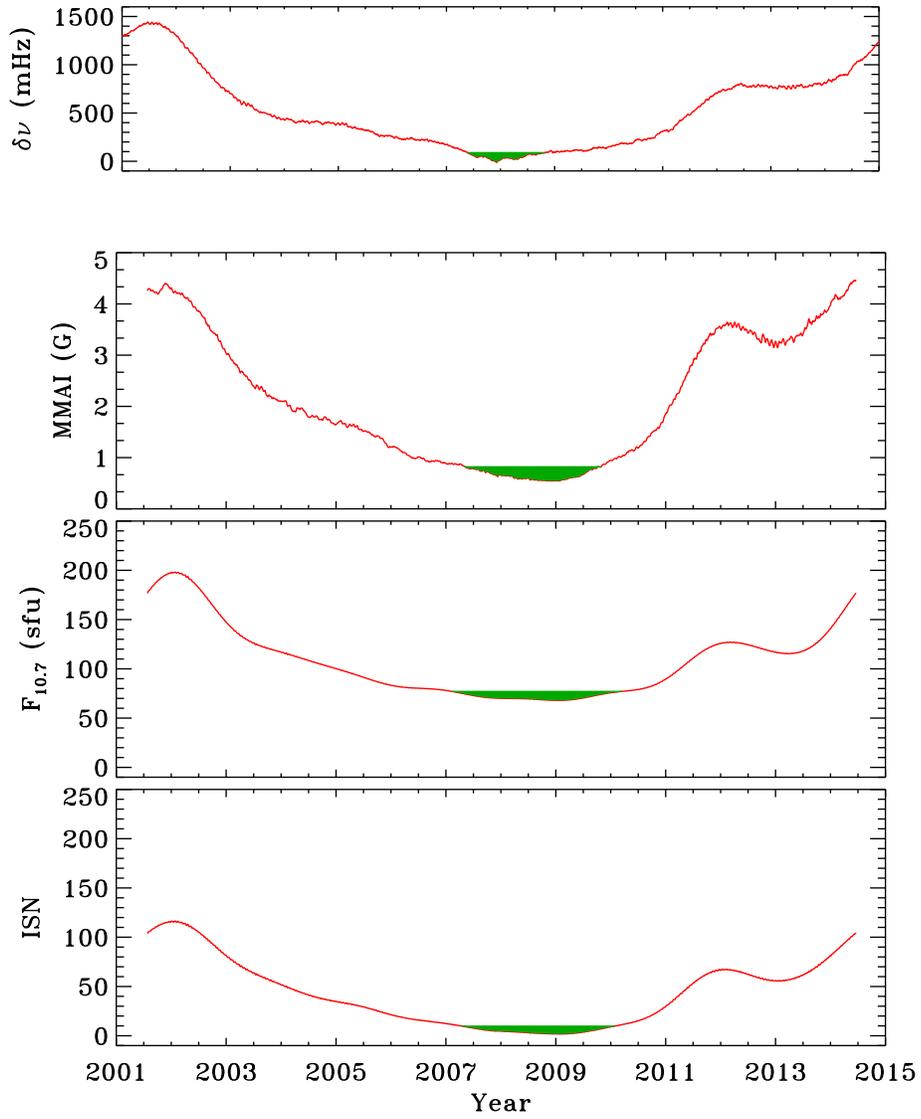}
   \caption{Temporal evolution of (a) $\delta\nu$, (b) MMAI, (c) F$_{\rm 10.7}$  and (d) ISN showing the extend of the solar minimum phase 
(filled area) which are overlaid on the running mean over 181 points.}
   \label{F-fig12}
   \end{figure*}

In order to better characterize and quantify the differences between the frequency shifts and activity proxies during the minimum phase, we calculate the length of the minimum period by defining a quiet phase as:  $$(maximum(x)-minimum(x))*0.07+minimum(x),$$ where x represents either  frequency shifts or activity proxies.  The factor of 0.07 is chosen based on the observed quiet phase in ISN and $F_{10.7}$. The temporal evolution of $\delta\nu$, MMAI, $F_{10.7}$ and ISN showing the extent of the minimum phase by filled green color is illustrated in Figure~\ref{F-fig12}.  It is evident that the minimum phase in $\delta\nu$ lasted for a brief period compared to the activity.  The minimum and maximum values of these parameters along with the start time, end time and the length of the quiet phase are listed in Table~\ref{T-tab4}. The minimum phase in frequency shifts lasted for a period of about 16 months while all the activity proxies lasted for a period of about 30 months or longer.  We have also computed these parameters separately for the northern and southern hemispheres. The values are listed in Table~\ref{T-tab4}. Confirming to known results, the activity (and hence the shifts) are found to be stronger in the southern hemisphere as compared to the northern hemisphere. We also note that the minimum phase lasted longer in northern hemisphere compared to the southern hemisphere,  probably, an indirect manifestation of the stronger activity in the southern hemisphere. We also find an exception in the northern hemisphere where the length of the quiet phase seen in $\delta\nu$ lasted slightly longer as compared to the magnetic activity.

 We further computed the length of the quiet phase for each latitude. The starting and ending epoch and the length of the minimum phase 
for $\delta\nu$ and MAI are listed in Table~\ref{T-tab5}; values above $\pm$~37.5\arcdeg\ latitudes are omitted since reliable measurements could not be obtained probably due to small variations between the minimum and maximum phase. Confirming to the earlier findings, the span of the quiet phase seen in frequency shifts is found to be reduced as compared to the magnetic activity. 

 \begin{deluxetable*}{lrrcccc}
\tablewidth{0pt}
\tabletypesize{\scriptsize} 
\tablecaption{Minimum and Maximum Values of Different Parameters, Start Time, End Time and Length of the Quiet Phase. \label{T-tab4}}
\tablehead{
\colhead{Index}& \colhead{Min}& \colhead{Max}&\colhead{Start}&\colhead{End}&\colhead{Length}}
\startdata
&&&All Data &\\
\cline{4-4}
&&&&&\\
$\delta\nu$ &100.2&1455.4&2007 Jun&2008 Oct&17\\
MAI &0.8&5.0&2007 May&2009 Oct&30\\
$F_{10.7}$ &82.8&241.7&2007 Feb&2010 Mar&38\\
ISN&0.0&159.5&2007 Apr&2010 Jan&34\\
&&&&&\\
&&&Northern Hemisphere&\\
\cline{4-4}
&&&&&\\
$\delta\nu$ &180.8&1324.4&2007 Jul&2010 Sep&39\\
MAI &0.9&4.8&2006 Aug&2009 Jun&35\\
HSN&0.0&80.5&2006 Mar&2009 Sep&43\\
&&&&&\\
&&&Southern Hemisphere&\\
\cline{4-4}
&&&&&\\
$\delta\nu$ &28.7&1700.0&2007 Jun&2008 Oct&17\\
MAI &0.7&5.6&2008 Mar&2010 Feb&24\\
HSN&0.0&92.9&2007 Sep&2010 Jun&34\\
\enddata
\tablecomments{Units of $\delta\nu$\ is nHz, MAI is Gauss and $F{10.7}$ is  sfu. The length is measured in month. }
\end{deluxetable*}

\begin{deluxetable*}{rllccllc}
\tablewidth{0pt}
\tabletypesize{\scriptsize} 
\tablecaption{Start Time, End Time and Length of the Quiet Phase at Various Latitudes.  \label{T-tab5}}
\tablehead{
\colhead{Latitude}&\multicolumn{3}{c}{MAI}&\colhead{}&\multicolumn{3}{c}{$\delta\nu$}\\
\cline{2-4} \cline{6-8}
&&&&&&&\\
\colhead{}&\colhead{Start}&\colhead{End}&\colhead{Length}&&\colhead{Start}&\colhead{End}&\colhead{Length\tablenotemark{a}}}
\startdata
0.0\arcdeg&2008 Jun&2011 Apr&35&&2009 Mar&2011 Dec&34\\
7.5\arcdeg&2006 Apr&2010 Jun&51&&2006 Dec&2010 Sep&48\\
15.0\arcdeg&2006 Feb&2009 Sep&44&&2006 Apr&2009 Sep&42\\
22.5\arcdeg&2006 Jan&2009 Jan&37&&2006 Feb&2008 Jul&30\\
30.0\arcdeg&2006 Mar&2008 Sep&31&&2005 Oct&2008 Mar&30\\
37.5\arcdeg&2007 Jan&2009 Sep&33&&2005 Aug&2008 Jan&30\\
-7.5\arcdeg&2008 Aug&2011 Jun&35&&2009 Jan&2011 Sep&33\\
-15.0\arcdeg&2008 Mar&2010 Jul&29&&2008 Mar&2010 Jul&29\\
-22.5\arcdeg&2006 Apr&2009 Oct&41&&2006 Oct&2010 Jan&40\\
-30.0\arcdeg&2006 Apr&2009 Apr&37&&2006 Jan&2008 Dec&36\\
-37.5\arcdeg&2007 Mar&2009 May&27&&2007 Jan&2008 Dec&24\\
\enddata
\tablenotetext{a}{Length is measured in month. }
\end{deluxetable*}

\section{Conclusions} 
In this work we have analyzed high-degree mode frequencies of the Sun over a period  spanning about 13 years to understand their 
behavior with the solar cycle.  These frequencies are obtained from the application of ring-diagram technique to the high-resolution Doppler observations made by the Global Oscillation Network Group and covers the period of 2001 July 26 to 2014 June 19.  

We compared the mean frequency shifts with three activity proxies-- the 10.7 cm radio flux, the International Sunspot Number and the 
strength of the local magnetic field measured by the magnetic activity index. In general, a good agreement 
is found between the shifts and activity indices and the correlation coefficients are found to be comparable with intermediate degree mode frequencies.  Although we do not have data covering all of cycle 23 or cycle 24, the phase-wise variations of the
shifts indicated that the frequencies responded to  solar activity equally well in both the cycles although the shifts per unit activity hinted that the cycle 24 is weaker than cycle 23. We also noticed that the frequency shifts do not show a prolonged  minimum phase as is seen 
in many activity indicators suggesting weak correlations between the shifts and activity indices during the activity minimum period. 

For the first time, we also computed frequency shifts over the north and south hemispheres separately and analyzed them as  a function of hemispheric sunspot number and magnetic activity index. The asymmetric nature of the magnetic activity is also reflected in the frequency shifts. We note that 
the shifts behaved similarly in the  northern and southern hemisphere until the end of 2007 and diverged thereafter. This indicates that the Sun had started showing deviations even before the onset of the minimum phase and that the mode frequencies sensed these changes before it was manifested otherwise at the visible surface. Further, we find that the shifts reached their maximum amplitude at two different periods--the amplitude of shifts in northern hemisphere peaked  during late 2011 more than two years earlier than the south.  Such asymmetric behavior 
between the ascending and descending phase of solar cycle 23 has been studied by \citet{sct-sim13} and it is argued that  
an increase of asymmetry might be used as a precursor of a grand minima. Theoretical calculations also indicate that strong asymmetries are induced by the quadrupolar component and characterize periods before entering or leaving the grand minima state \citep{sct-tobias96, sct-knob98}.  
Thus in our view the asymmetry associated with the frequency shifts may have certain implications for the  dynamo action taking place either in the sub-surface shear layer or at the base of the convection zone.

Since the shifts and the hemispheric activity indices are found to be significantly correlated, we suggest that the frequency shifts may be used as an indicator of  hemispheric activity  since not many indices are measured over the two hemispheres separately.  We also investigate the variations at different latitudes and conclude that the shifts in active latitudes correlate well with the local magnetic activity index.  We believe that this analysis provides impetus to  analyze solar cycle 24 in terms of hemispheric asymmetry and different phases to comprehend the magnetic cycle of the Sun.  

\acknowledgments
 We thank the referee for comments that have improved the paper. We also thank John Leibacher and Shukur Kholikov for useful discussions. 
 This work utilizes data obtained by the NSO Integrated Synoptic Program (NISP), managed by the National Solar Observatory, which
is operated by AURA, Inc. under a cooperative agreement with the
National Science Foundation. The data were acquired by instruments
operated by the Big Bear Solar Observatory, High Altitude Observatory,
Learmonth Solar Observatory, Udaipur Solar Observatory, Instituto de
Astrof\'{\i}sico de Canarias, and Cerro Tololo Interamerican
Observatory. This study also includes data from
SOHO/MDI.  SOHO is a mission of international cooperation between ESA
and NASA. This work has made use of data 
available at NOAA's National Geophysical Data Center (NGDC). We thank S. Korzennik for providing us with the high-degree GONG data. 
This work is partially supported by NASA grant NNH12AT11I. A part of this work  was performed  under the auspices of the SPACEINN Framework of the European Union (EU FP7). 

\bibliography{acoustic}

\begin{thebibliography}{}
\expandafter\ifx\csname natexlab\endcsname\relax\def\natexlab#1{#1}\fi

\bibitem[{{Baldner} \& {Basu}(2008)}]{sct-baldner08}
{Baldner}, C.~S., \& {Basu}, S. 2008, \apj, 686, 1349

\bibitem[{{Basu} {et~al.}(2004){Basu}, {Antia}, \& {Bogart}}]{sct-basu04}
{Basu}, S., {Antia}, H.~M., \& {Bogart}, R.~S. 2004, \apj, 610, 1157

\bibitem[{{Basu} {et~al.}(2012){Basu}, {Broomhall}, {Chaplin}, \&
  {Elsworth}}]{sct-basu12}
{Basu}, S., {Broomhall}, A.-M., {Chaplin}, W.~J., \& {Elsworth}, Y. 2012, \apj,
  758, 43

\bibitem[{{Bogart} {et~al.}(2002){Bogart}, {Basu}, \& {Antia}}]{sct-rick02}
{Bogart}, R.~S., {Basu}, S., \& {Antia}, H.~M. 2002, in ESA Special
  Publication, Vol. 508, From Solar Min to Max: Half a Solar Cycle with SOHO,
  ed. A.~{Wilson}, 145--148

\bibitem[{{Broomhall} {et~al.}(2009){Broomhall}, {Chaplin}, {Elsworth},
  {Fletcher}, \& {New}}]{sct-bison09}
{Broomhall}, A.-M., {Chaplin}, W.~J., {Elsworth}, Y., {Fletcher}, S.~T., \&
  {New}, R. 2009, \apjl, 700, L162

\bibitem[{{Broomhall} {et~al.}(2012){Broomhall}, {Chaplin}, {Elsworth}, \&
  {Simoniello}}]{sct-anne12}
{Broomhall}, A.-M., {Chaplin}, W.~J., {Elsworth}, Y., \& {Simoniello}, R. 2012,
  \mnras, 420, 1405

\bibitem[{{Chaplin} {et~al.}(2001){Chaplin}, {Appourchaux}, {Elsworth},
  {Isaak}, \& {New}}]{sct-chap01}
{Chaplin}, W.~J., {Appourchaux}, T., {Elsworth}, Y., {Isaak}, G.~R., \& {New},
  R. 2001, \mnras, 324, 910

\bibitem[{{Chaplin} {et~al.}(2007){Chaplin}, {Elsworth}, {Miller}, {Verner}, \&
  {New}}]{sct-chap07}
{Chaplin}, W.~J., {Elsworth}, Y., {Miller}, B.~A., {Verner}, G.~A., \& {New},
  R. 2007, \apj, 659, 1749

\bibitem[{{Chou} \& {Serebryanskiy}(2005)}]{sct-chou05}
{Chou}, D.-Y., \& {Serebryanskiy}, A. 2005, \apj, 624, 420

\bibitem[{{Corbard} {et~al.}(2003){Corbard}, {Toner}, {Hill}, {Hanna}, {Haber},
  {Hindman}, \& {Bogart}}]{sct-corbard03}
{Corbard}, T., {Toner}, C., {Hill}, F., {et~al.} 2003, in ESA Special
  Publication, Vol. 517, GONG+ 2002. Local and Global Helioseismology: the
  Present and Future, ed. H.~{Sawaya-Lacoste}, 255--258

\bibitem[{{Dziembowski} \& {Goode}(2005)}]{sct-dzi05}
{Dziembowski}, W.~A., \& {Goode}, P.~R. 2005, \apj, 625, 548

\bibitem[{{Elsworth} {et~al.}(1994){Elsworth}, {Howe}, {Isaak}, {McLeod},
  {Miller}, {New}, {Speake}, \& {Wheeler}}]{sct-els94}
{Elsworth}, Y., {Howe}, R., {Isaak}, G.~R., {et~al.} 1994, \apj, 434, 801

\bibitem[{{Elsworth} {et~al.}(1990){Elsworth}, {Howe}, {Isaak}, {McLeod}, \&
  {New}}]{sct-els90}
{Elsworth}, Y., {Howe}, R., {Isaak}, G.~R., {McLeod}, C.~P., \& {New}, R. 1990,
  \nat, 345, 322

\bibitem[{{Foullon} \& {Roberts}(2005)}]{sct-foullon05}
{Foullon}, C., \& {Roberts}, B. 2005, \aap, 439, 713

\bibitem[{{Haber} {et~al.}(2000){Haber}, {Hindman}, {Toomre}, {Bogart},
  {Thompson}, \& {Hill}}]{sct-haber00}
{Haber}, D.~A., {Hindman}, B.~W., {Toomre}, J., {et~al.} 2000, \solphys, 192,
  335

\bibitem[{{Hill}(1988)}]{sct-hill88}
{Hill}, F. 1988, \apj, 333, 996

\bibitem[{{Hindman} {et~al.}(2000){Hindman}, {Haber}, {Toomre}, \&
  {Bogart}}]{sct-hindman00}
{Hindman}, B., {Haber}, D., {Toomre}, J., \& {Bogart}, R. 2000, \solphys, 192,
  363

\bibitem[{{Howe} {et~al.}(1999){Howe}, {Komm}, \& {Hill}}]{sct-howe99}
{Howe}, R., {Komm}, R., \& {Hill}, F. 1999, \apj, 524, 1084

\bibitem[{{Howe} {et~al.}(2002){Howe}, {Komm}, \& {Hill}}]{sct-howe02}
{Howe}, R., {Komm}, R.~W., \& {Hill}, F. 2002, \apj, 580, 1172

\bibitem[{{Howe} {et~al.}(2004){Howe}, {Komm}, {Hill}, {Haber}, \&
  {Hindman}}]{sct-rhowe04}
{Howe}, R., {Komm}, R.~W., {Hill}, F., {Haber}, D.~A., \& {Hindman}, B.~W.
  2004, \apj, 608, 562

\bibitem[{{Jain} {et~al.}(2009){Jain}, {Tripathy}, \& {Hill}}]{sct-jain09}
{Jain}, K., {Tripathy}, S.~C., \& {Hill}, F. 2009, \apj, 695, 1567

\bibitem[{{Jain} {et~al.}(2011){Jain}, {Tripathy}, \& {Hill}}]{sct-jain11}
---. 2011, \apj, 739, 6

\bibitem[{{Jain} \& {Roberts}(1996)}]{sct-rjain96}
{Jain}, R., \& {Roberts}, B. 1996, \apj, 456, 399

\bibitem[{{Knobloch} {et~al.}(1998){Knobloch}, {Tobias}, \&
  {Weiss}}]{sct-knob98}
{Knobloch}, E., {Tobias}, S.~M., \& {Weiss}, N.~O. 1998, \mnras, 297, 1123

\bibitem[{{Korzennik} {et~al.}(2013{\natexlab{a}}){Korzennik}, {Eff-Darwich},
  {Larson}, {Rabello-Soares}, \& {Schou}}]{sct-kornso50}
{Korzennik}, S.~G., {Eff-Darwich}, A., {Larson}, T.~P., {Rabello-Soares},
  M.~C., \& {Schou}, J. 2013{\natexlab{a}}, in Astronomical Society of the
  Pacific Conference Series, Vol. 478, Fifty Years of Seismology of the Sun and
  Stars, ed. K.~{Jain}, S.~C. {Tripathy}, F.~{Hill}, J.~W. {Leibacher}, \&
  A.~A. {Pevtsov}, 173

\bibitem[{{Korzennik} {et~al.}(2013{\natexlab{b}}){Korzennik},
  {Rabello-Soares}, {Schou}, \& {Larson}}]{sct-kor2013}
{Korzennik}, S.~G., {Rabello-Soares}, M.~C., {Schou}, J., \& {Larson}, T.~P.
  2013{\natexlab{b}}, \apj, 772, 87

\bibitem[{{Kuhn}(1988)}]{sct-kuhn88}
{Kuhn}, J.~R. 1988, \apjl, 331, L131

\bibitem[{{Norton} {et~al.}(2014){Norton}, {Charbonneau}, \&
  {Passos}}]{sct-aanorton2014}
{Norton}, A.~A., {Charbonneau}, P., \& {Passos}, D. 2014, \ssr, 186, 251

\bibitem[{{Rabello-Soares}(2011)}]{sct-crs11}
{Rabello-Soares}, M.~C. 2011, Journal of Physics Conference Series, 271, 012026

\bibitem[{{Rabello-Soares} {et~al.}(2008{\natexlab{a}}){Rabello-Soares},
  {Bogart}, \& {Basu}}]{sct-rab08}
{Rabello-Soares}, M.~C., {Bogart}, R.~S., \& {Basu}, S. 2008{\natexlab{a}},
  Journal of Physics Conference Series, 118, 012084

\bibitem[{{Rabello-Soares} {et~al.}(2008{\natexlab{b}}){Rabello-Soares},
  {Korzennik}, \& {Schou}}]{sct-crs08}
{Rabello-Soares}, M.~C., {Korzennik}, S.~G., \& {Schou}, J. 2008{\natexlab{b}},
  Advances in Space Research, 41, 861

\bibitem[{{Rajaguru} {et~al.}(2001){Rajaguru}, {Basu}, \&
  {Antia}}]{sct-rajguru01}
{Rajaguru}, S.~P., {Basu}, S., \& {Antia}, H.~M. 2001, \apj, 563, 410

\bibitem[{{Rhodes} {et~al.}(2002){Rhodes}, {Reiter}, \& {Schou}}]{sct-rhodes02}
{Rhodes}, Jr., E.~J., {Reiter}, J., \& {Schou}, J. 2002, in ESA Special
  Publication, Vol. 508, From Solar Min to Max: Half a Solar Cycle with SOHO,
  ed. A.~{Wilson}, 37--40

\bibitem[{{Rhodes} {et~al.}(2011){Rhodes}, {Reiter}, {Schou}, {Larson},
  {Scherrer}, {Brooks}, {McFaddin}, {Miller}, {Rodriguez}, \&
  {Yoo}}]{sct-rhodes11}
{Rhodes}, Jr., E.~J., {Reiter}, J., {Schou}, J., {et~al.} 2011, Journal of
  Physics Conference Series, 271, 012029

\bibitem[{{Riley} {et~al.}(2014){Riley}, {Ben-Nun}, {Linker}, {Mikic},
  {Svalgaard}, {Harvey}, {Bertello}, {Hoeksema}, {Liu}, \&
  {Ulrich}}]{sct-harvey14}
{Riley}, P., {Ben-Nun}, M., {Linker}, J.~A., {et~al.} 2014, \solphys, 289, 769

\bibitem[{{Rose} {et~al.}(2003){Rose}, {Rhodes}, {Reiter}, \&
  {Rudnisky}}]{sct-rhodes03}
{Rose}, P., {Rhodes}, Jr., E.~J., {Reiter}, J., \& {Rudnisky}, W. 2003, in ESA
  Special Publication, Vol. 517, GONG+ 2002. Local and Global Helioseismology:
  the Present and Future, ed. H.~{Sawaya-Lacoste}, 373--376

\bibitem[{{Salabert} {et~al.}(2009){Salabert}, {Garc{\'{\i}}a}, {Pall{\'e}}, \&
  {Jim{\'e}nez-Reyes}}]{sct-salabert09}
{Salabert}, D., {Garc{\'{\i}}a}, R.~A., {Pall{\'e}}, P.~L., \&
  {Jim{\'e}nez-Reyes}, S.~J. 2009, \aap, 504, L1

\bibitem[{{Salabert} {et~al.}(2015){Salabert}, {Garc{\'{\i}}a}, \&
  {Turck-Chi{\`e}ze}}]{sct-david15}
{Salabert}, D., {Garc{\'{\i}}a}, R.~A., \& {Turck-Chi{\`e}ze}, S. 2015, \aap,
  578, A137

\bibitem[{{Schou}(1999)}]{sct-schou99}
{Schou}, J. 1999, \apjl, 523, L181

\bibitem[{{Simoniello} {et~al.}(2013{\natexlab{a}}){Simoniello}, {Jain},
  {Tripathy}, {Baldner}, {Turck-Chi{\`e}ze}, \& {Hill}}]{sct-sim13}
{Simoniello}, R., {Jain}, K., {Tripathy}, S.~C., {et~al.} 2013{\natexlab{a}},
  in Astronomical Society of the Pacific Conference Series, Vol. 478, Fifty
  Years of Seismology of the Sun and Stars, ed. K.~{Jain}, S.~C. {Tripathy},
  F.~{Hill}, J.~W. {Leibacher}, \& A.~A. {Pevtsov}, 167

\bibitem[{{Simoniello} {et~al.}(2013{\natexlab{b}}){Simoniello}, {Jain},
  {Tripathy}, {Turck-Chi{\`e}ze}, {Baldner}, {Finsterle}, {Hill}, \&
  {Roth}}]{sct-rosy13}
{Simoniello}, R., {Jain}, K., {Tripathy}, S.~C., {et~al.} 2013{\natexlab{b}},
  \apj, 765, 100

\bibitem[{{Tobias}(1996)}]{sct-tobias96}
{Tobias}, S.~M. 1996, \aap, 307, L21

\bibitem[{{Tripathy} {et~al.}(2007){Tripathy}, {Hill}, {Jain}, \&
  {Leibacher}}]{sct-sol07}
{Tripathy}, S.~C., {Hill}, F., {Jain}, K., \& {Leibacher}, J.~W. 2007,
  \solphys, 243, 105

\bibitem[{{Tripathy} {et~al.}(2010{\natexlab{a}}){Tripathy}, {Jain}, \&
  {Hill}}]{sct10iia}
{Tripathy}, S.~C., {Jain}, K., \& {Hill}, F. 2010{\natexlab{a}}, in Magnetic
  Coupling between the Interior and Atmosphere of the Sun, ed. S.~S. {Hasan} \&
  R.~J. {Rutten}, 374--377

\bibitem[{{Tripathy} {et~al.}(2011){Tripathy}, {Jain}, \& {Hill}}]{sct11}
{Tripathy}, S.~C., {Jain}, K., \& {Hill}, F. 2011, Journal of Physics
  Conference Series, 271, 012024

\bibitem[{{Tripathy} {et~al.}(2013{\natexlab{a}}){Tripathy}, {Jain}, \&
  {Hill}}]{sct-sol13}
---. 2013{\natexlab{a}}, \solphys, 282, 1, paper I

\bibitem[{{Tripathy} {et~al.}(2013{\natexlab{b}}){Tripathy}, {Jain}, \&
  {Hill}}]{sct-nso50}
{Tripathy}, S.~C., {Jain}, K., \& {Hill}, F. 2013{\natexlab{b}}, in ASP Conf.
  Ser., Vol. 478, Fifty Years of Seismology of the Sun and Stars, ed.
  K.~{Jain}, S.~C. {Tripathy}, F.~{Hill}, J.~{Leibacher}, \& A.~A. {Pevtsov},
  221

\bibitem[{{Tripathy} {et~al.}(2010{\natexlab{b}}){Tripathy}, {Jain}, {Hill}, \&
  {Leibacher}}]{sct-2010apj}
{Tripathy}, S.~C., {Jain}, K., {Hill}, F., \& {Leibacher}, J.~W.
  2010{\natexlab{b}}, \apjl, 711, L84

\bibitem[{{Woodard} {et~al.}(1991){Woodard}, {Libbrecht}, {Kuhn}, \&
  {Murray}}]{sct-woodard91}
{Woodard}, M.~F., {Libbrecht}, K.~G., {Kuhn}, J.~R., \& {Murray}, N. 1991,
  \apjl, 373, L81

\bibitem[{{Woodard} \& {Noyes}(1985)}]{sct-woodard85}
{Woodard}, M.~F., \& {Noyes}, R.~W. 1985, \nat, 318, 449

\end{thebibliography}
\end{document}